\documentclass[twocolumn,prl,showpacs]{revtex4}
\usepackage{graphics}

\begin{document}

\title{Quantum locking of mirrors in interferometers}

\author{Jean-Michel Courty}
\email{courty@spectro.jussieu.fr}
\author{Antoine Heidmann}
\email{heidmann@spectro.jussieu.fr}
\author{Michel Pinard}
\email{pinard@spectro.jussieu.fr}

\affiliation{Laboratoire Kastler Brossel, Case 74, 4 place Jussieu, F75252
Paris Cedex 05, France}
\thanks{Unit\'{e} mixte de recherche du Centre National de la Recherche
Scientifique, de l'Ecole Normale Sup\'{e}rieure et de l'Universit\'{e} Pierre
et Marie Curie} \homepage{www.spectro.jussieu.fr/Mesure}

\date{January 6, 2003}

\begin{abstract}
We show that quantum noise in very sensitive interferometric measurements
such as gravitational-waves detectors can be drastically modified by quantum
feedback.\ We present a new scheme based on active control to lock the motion
of a mirror to a reference mirror at the quantum level.\ This simple
technique allows one to reduce quantum effects of radiation pressure and to
greatly enhance the sensitivity of the detection.
\end{abstract}

\pacs{42.50.Lc, 03.65.Ta, 04.80.Nn}

\maketitle

Quantum noise of light plays an important role in the sensitivity limits of
optical measurements such as gravitational-wave interferometers. A
gravitational wave is detected as a phase difference between the two optical
arms of a Michelson interferometer\cite{Bradaschia90,Abramovici92}. The
detection is limited by two fundamental noises, the phase noise of the laser
beam which leads to an error in the measurement of the arm's length, and the
intensity noise which induces displacements of the mirrors via radiation
pressure. Both noises are conjugate and the sensitivity at a given frequency
can be optimized by choosing the light intensity so that phase and
radiation-pressure noises are equal\cite{Caves81,Braginsky92}.

Since the mechanical response of a mirror depends on frequency, its
sensitivity to radiation pressure is also frequency dependent. For a
suspended mirror the mechanical response decreases with frequency and
radiation-pressure noise is dominant at low frequency, whereas phase noise is
dominant at high frequency. Optimization of both noises is then obtained at
only one frequency. This behavior can be changed by using non classical
states of light\cite{Braginsky92,Jaekel90,McKenzie02} at the expense of a
larger complexity of the system in order to generate and to adapt squeezed
states to the interferometer\cite{Kimble02}.

We present in this paper a new technique to increase the sensitivity in
interferometers, based on a reduction of radiation-pressure noise by active
control of the mirror motion. This alternative approach does not rely on the
use of non classical state of light and has little impact on the system
complexity.

Active control can indeed efficiently reduce classical noise, as, for
example, thermal fluctuations in cold-damped mechanical systems
\cite{Milatz53,Grassia00}, but also the quantum back-action noise of a
measurement\cite{Wiseman95}. A cold-damping technique has been implemented in
optomechanical systems to reduce the thermal motion of a mirror\cite
{Mancini98,Cohadon99,Pinard00}, and it can in principle be used in a quantum
regime to cool the mirror temperature down to zero\cite{Courty01}.

We show that a feedback control can lock a mirror with respect to the
position of a less noisy reference mirror, thus reducing the
radiation-pressure noise in the interferometer. This quantum locking
increases the sensitivity bandwidth at low frequency without degrading
performances at high frequency where phase noise is dominant.

\begin{figure}
{\resizebox{7 cm}{!}{\includegraphics{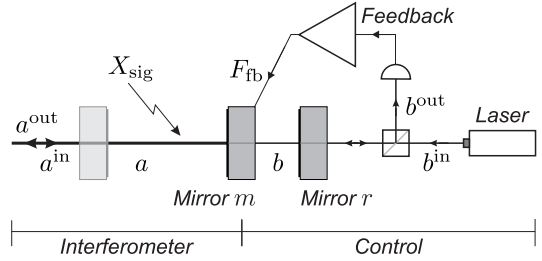}}}
\caption{A Fabry-Perot
cavity and a light field $a$ are used to measure a signal $X_{\rm sig}$
corresponding to a variation of the cavity length.\ The end mirror $m$ is
actively controlled by a feedback loop via a force $F_{\rm fb}$ delivered by
an optomechanical displacement sensor (control cavity made of mirrors $m$ and
$r$, field $b$).} \label{Fig_Scheme}
\end{figure}

The basic setup is shown in Fig. \ref{Fig_Scheme}.\ The interferometric
measurement is made by a single Fabry-Perot cavity which can be considered as
one arm of a gravitational-wave detector. The signal induces a variation
$X_{{\rm sig}}$ of the cavity length which is detected by sending a light
field $a$ in the cavity. We focus on the active control of a single mirror of
the cavity, namely, the end mirror $m$ in Fig. \ref{Fig_Scheme}. The mirror
motion is measured by a second light beam $b$ interacting with a short cavity
made of the mirror $m$ and a reference mirror $r.$ The result of the
measurement is used to apply a correcting force on the mirror.

The ultimate sensitivity of a gravitational-wave interferometer is determined
only by quantum noise of light. Classical noises such as seismic or thermal
noises can in principle be eliminated and will not be considered in the
following. Quantum fluctuations of field $a$ at frequency $\Omega $ are
described by the quantum annihilation operator $a\left[ \Omega \right] $,
whereas the mean field is given by a complex amplitude $\alpha $ normalized
in such a way that $\left| \alpha \right| ^{2}$ corresponds to a photon flux
\cite{Reynaud92}. For a lossless single-ended cavity resonant with the laser
field, the incident, intracavity, and output mean field amplitudes can be
taken as real.\ It is then convenient to describe quantum fluctuations by the
amplitude and phase quadratures $a_{1}$ and $a_{2}$ corresponding to the real
and imaginary parts of the operator $a$,
\begin{eqnarray}
a_{1}\left[ \Omega \right] &=&a\left[ \Omega \right] +a^{\dagger }\left[
\Omega \right] , \\
a_{2}\left[ \Omega \right] &=&-i\left( a\left[ \Omega \right] -a^{\dagger }%
\left[ \Omega \right] \right) .
\end{eqnarray}
Noise spectra of incident quadratures $a_{1}^{{\rm in}}$ and $a_{2}^{{\rm in}%
}$ for an incident coherent state are
\begin{equation}
\sigma _{a_{1}a_{1}}^{{\rm in}}=\sigma _{a_{2}a_{2}}^{{\rm in}}=1,\quad
\sigma _{a_{1}a_{2}}^{{\rm in}}=0,
\end{equation}
where $\sigma _{a_{i}a_{j}}^{{\rm in}}$ is defined as the quantum average of
the symmetrized product of operators $a_{i}^{{\rm in}}$ and $a_{j}^{{\rm in}%
} $,
\begin{equation}
\left\langle a_{i}^{{\rm in}}\left[ \Omega \right] \cdot a_{j}^{{\rm in}}%
\left[ \Omega ^{\prime }\right] \right\rangle =2\pi \delta \left( \Omega
+\Omega ^{\prime }\right) \sigma _{a_{i}a_{j}}^{{\rm in}}\left[ \Omega %
\right] .
\end{equation}
For frequency $\Omega $ smaller than the cavity bandwidth, the amplitude
quadrature is left unchanged by the cavity ($a_{1}^{{\rm out}}=a_{1}^{{\rm in%
}}$), whereas the input-output phase shift is proportional to the variation $%
X_{{\rm m}}+X_{{\rm sig}}$ of the cavity length ($X_{{\rm m}}$ is the
displacement of mirror $m$)\cite{Courty01},

\begin{equation}
a_{2}^{{\rm out}}=a_{2}^{{\rm in}}+2\xi _{{\rm a}}\left( X_{{\rm m}}+X_{{\rm %
sig}}\right) .  \label{Equ_a2out}
\end{equation}
The optomechanical coupling parameter $\xi _{{\rm a}}$ is related to the
intracavity mean field amplitude $\alpha $, to the cavity finesse ${\cal F}_{%
{\rm a}}$ and to the wavevector $k_{0}$ of the light,
\begin{equation}
\xi _{{\rm a}}=2k_{0}\alpha \sqrt{2{\cal F}_{{\rm a}}/\pi }.  \label{Equ_Xia}
\end{equation}

The interferometric measurement provides an estimator $\hat{X}_{{\rm sig}}$
of the signal obtained by a normalization of the output phase $a_{2}^{{\rm %
out}}$ as a displacement [eq. (\ref{Equ_a2out})]. $\hat{X}_{{\rm sig}}$ is
the sum of the signal $X_{{\rm sig}}$ and extra noise terms,
\begin{equation}
\hat{X}_{{\rm sig}}=\frac{1}{2\xi _{{\rm a}}}a_{2}^{{\rm out}}=X_{{\rm sig}}+%
\frac{1}{2\xi _{{\rm a}}}a_{2}^{{\rm in}}+X_{{\rm m}}.  \label{Equ_Xsig}
\end{equation}
The first noise term is related to the incident phase noise $a_{2}^{{\rm
in}}$, and the second corresponds to unwanted variations of the cavity length
through the motion of mirror $m$.

The determination of the interferometer sensitivity thus requires one to know
the motion of mirror $m$. For a free interferometer (no control cavity and no
feedback) the evolution of the velocity $V_{{\rm m}}=-i\Omega X_{{\rm m}}$ is
governed by the radiation pressure of the intracavity field $a$ whose
fluctuations can be expressed in terms of the incident amplitude quadrature $%
a_{1}^{{\rm in}}$,
\begin{equation}
Z_{{\rm m}}V_{{\rm m}}=\hbar \xi _{{\rm a}}a_{1}^{{\rm in}},
\label{Equ_Vmfree}
\end{equation}
where $Z_{{\rm m}}$ is the mechanical impedance of mirror $m$. In the
frequency band relevant for a gravitational-wave interferometer, the
suspended mirror can be considered as a free mass with a mechanical impedance
$Z_{{\rm m}}$ related to the mirror mass $M_{{\rm m}}$ by
\begin{equation}
Z_{{\rm m}}\simeq -i\Omega M_{{\rm m}}.
\end{equation}

\begin{figure}
{\resizebox{6 cm}{!}{\includegraphics{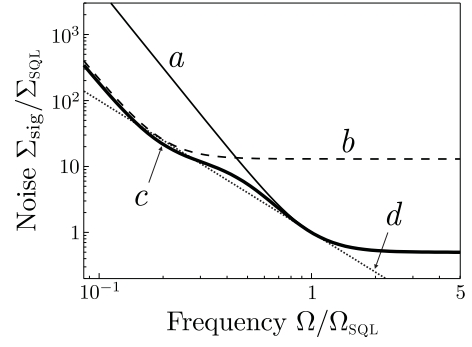}}} \caption{Equivalent input
noise $\Sigma _{{\rm sig}}$ in the interferometric measurement as a function
of frequency $\Omega $ for the free interferometer (curve {\it a}), for an
infinite feedback gain (curve {\it b}), and for an optimum gain (curve {\it
c}). Curve {\it d} is the standard quantum limit. Frequency is normalized to
the {\sc sql} frequency $\Omega _{\text{{\sc sql}}}$ and noise to $\Sigma
_{\text{{\sc sql}}}$. Optomechanical coupling in the control cavity is $\xi
_{{\rm b}}=\xi _{{\rm a}}/5$.} \label{Fig_Sensit}
\end{figure}

The sensitivity of the measurement is described as an equivalent input noise
$\Sigma _{{\rm sig}}$ equal to the spectrum of noises added in the estimator
$\hat{X}_{{\rm sig}}$ [eq. (\ref{Equ_Xsig})]. For a coherent input light one
gets
\begin{equation}
\Sigma _{{\rm sig}}=\frac{1}{4\xi _{{\rm a}}^{2}}+\frac{\hbar ^{2}\xi _{{\rm %
a}}^{2}}{\left| \Omega \right| ^{2}\left| Z_{{\rm m}}\right| ^{2}}.
\label{Equ_Ssigfree}
\end{equation}
The first term is a measurement error due to phase noise, and the second term
corresponds to the mirror motion induced by radiation pressure. As shown in
curve {\it a} of Fig. \ref{Fig_Sensit}, phase noise is dominant at high
frequency with a flat frequency dependence, whereas radiation pressure is
dominant at low frequency with a $1/\Omega ^{4}$ dependence. To reach a good
sensitivity at high frequency one has to choose a large optomechanical
parameter $\xi _{{\rm a}}$, but the analysis bandwidth is reduced at low
frequency by the increase of radiation pressure.

For a given optomechanical coupling $\xi _{{\rm a}}$ the sensitivity is
optimized only at a frequency $\Omega _{\text{{\sc sql}}}$ where
contributions of both noises are equal. This optimization leads to the
so-called standard quantum limit ({\sc sql}) $\Sigma _{\text{{\sc sql}}}$
given by
\begin{eqnarray}
\Omega _{\text{{\sc sql}}}^{2} &=&\frac{2\hbar \xi _{{\rm a}}^{2}}{M_{{\rm m}%
}},  \label{Equ_Osql} \\
\Sigma _{\text{{\sc sql}}} &=&\frac{\hbar }{M_{{\rm m}}\Omega _{\text{{\sc %
sql}}}^{2}}.
\end{eqnarray}
$\Sigma _{\text{{\sc sql}}}$ corresponds to the minimum noise reachable at a
given frequency by a proper choice of the optomechanical coupling (curve
{\it d} of Fig. \ref{Fig_Sensit}).

The addition of a feedback loop allows one to suppress radiation-pressure
effects by freezing the motion of mirror $m$. The measurement with the
control cavity, of course, has its own quantum limit which has to be taken
into account in a quantum analysis of the controlled interferometer, as well
as the motion of the reference mirror $r$ induced by the radiation pressure
of beam $b$. The control cavity indeed measures the position $X_{{\rm m}}$ of
mirror $m$ with respect to the position $X_{{\rm r}}$ of the reference mirror
$r$. The phase shift $b_{2}^{{\rm out}}$ at the output of cavity $b$ is given
by
\begin{equation}
b_{2}^{{\rm out}}=b_{2}^{{\rm in}}+2\xi _{{\rm b}}\left( X_{{\rm r}}-X_{{\rm %
m}}\right) ,  \label{Equ_b2out}
\end{equation}
with an optomechanical parameter $\xi _{{\rm b}}$ for cavity $b$ defined in
the same way as $\xi _{{\rm a}}$ [eq. (\ref{Equ_Xia})].

The measurement of the output phase provides an estimator $\hat{V}_{{\rm m}}$
for the motion of mirror $m$, proportional to the output quadrature $b_{2}^{%
{\rm out}}$ and normalized as a velocity. We choose a feedback force applied
on mirror $m$ proportional to this estimator\cite{Courty01},
\begin{equation}
F_{{\rm fb}}=-Z_{{\rm fb}}\hat{V}_{{\rm m}}=-Z_{{\rm fb}}\frac{i\Omega }{%
2\xi _{{\rm b}}}b_{2}^{{\rm out}},  \label{Equ_Ffb}
\end{equation}
where $Z_{{\rm fb}}$ is the transfer function of the feedback loop.

The motion of mirror $m$ now depends on radiation pressures from both
cavities and on the feedback force, whereas the motion of mirror $r$ depends
only on the radiation pressure of field $b$,
\begin{eqnarray}
Z_{{\rm m}}V_{{\rm m}} &=&\hbar \xi _{{\rm a}}a_{1}^{{\rm in}}-\hbar \xi _{%
{\rm b}}b_{1}^{{\rm in}}+F_{{\rm fb}},  \label{Equ_Vm} \\
Z_{{\rm r}}V_{{\rm r}} &=&\hbar \xi _{{\rm b}}b_{1}^{{\rm in}},
\label{Equ_Vr}
\end{eqnarray}
where $Z_{{\rm r}}$ is the mechanical impedance of mirror $r$. The resulting
motion of mirror $m$ is obtained from Eqs. (\ref{Equ_b2out}) to (\ref{Equ_Vm}%
),
\begin{eqnarray}
\left( Z_{{\rm m}}+Z_{{\rm fb}}\right) V_{{\rm m}} &=&\hbar \xi _{{\rm a}%
}a_{1}^{{\rm in}}-\hbar \xi _{{\rm b}}b_{1}^{{\rm in}}  \nonumber \\
&&+Z_{{\rm fb}}\left( V_{{\rm r}}-\frac{i\Omega }{2\xi _{{\rm b}}}b_{2}^{%
{\rm in}}\right) .  \label{Equ_Vmfb}
\end{eqnarray}
Combined with Eqs. (\ref{Equ_Xsig}) and (\ref{Equ_Vr}) one gets the estimator
$\hat{X}_{{\rm sig}}$ for the interferometric measurement with feedback,
\begin{eqnarray}
\hat{X}_{{\rm sig}} &=&X_{{\rm sig}}+\frac{1}{2\xi _{{\rm a}}}a_{2}^{{\rm in}%
}+\frac{1}{2\xi _{{\rm b}}}\frac{Z_{{\rm fb}}}{Z_{{\rm m}}+Z_{{\rm fb}}}%
b_{2}^{{\rm in}}  \nonumber \\
&&+\frac{i\hbar }{\Omega \left( Z_{{\rm m}}+Z_{{\rm fb}}\right) }\left( \xi
_{{\rm a}}a_{1}^{{\rm in}}-\frac{Z_{{\rm r}}-Z_{{\rm fb}}}{Z_{{\rm r}}}\xi _{%
{\rm b}}b_{1}^{{\rm in}}\right) \label{Equ_Xsigfb}
\end{eqnarray}

The first effect of the control is to change the dynamics of mirror $m$ by
adding a feedback-induced impedance $Z_{{\rm fb}}$ to the free mechanical
impedance $Z_{{\rm m}}$ [compare left parts of eqs. (\ref{Equ_Vmfree}) and (%
\ref{Equ_Vmfb})]. The influence of radiation-pressure effects due to the
interferometer is modified both for the mirror velocity and the
signal estimator. The associated noise [term proportional to quadrature $%
a_{1}^{{\rm in}}$ in eq. (\ref{Equ_Xsigfb})] is reduced for a large feedback
gain, without changing the phase noise (term proportional to quadrature $%
a_{2}^{{\rm in}}$).

The second effect of the control is to add fluctuating forces to mirror $m$.
The second term in Eq. (\ref{Equ_Vmfb}) corresponds to the radiation pressure
of beam $b$ acting on mirror $m$. The last terms are associated with the
presence of the feedback loop and correspond to noises in the control
measurement, due to either the phase noise of beam $b$ (term proportional to
$b_{2}^{{\rm in}}$) or the motion of the reference mirror $r$. As a
consequence, the estimator $\hat{X}_{{\rm sig}}$ exhibits additional
fluctuations related to the incident quadratures $b_{1}^{{\rm in}}$ and $%
b_{2}^{{\rm in}}$ of beam $b$.

We show now that it is possible to adapt both the optomechanical coupling $%
\xi _{{\rm b}}$ of the control cavity and the feedback gain $Z_{{\rm fb}}$ in
order to obtain a large increase of the interferometer sensitivity. The
optimization relies on a complete elimination of the radiation-pressure noise
due to the interferometer.

The underlying mechanism can be understood in the limit of a very large
feedback gain ($Z_{{\rm fb}}\rightarrow \infty $) in which case the error
signal $b_{2}^{{\rm out}}$ of the feedback loop is reduced to zero.\ Apart
from the phase noise of beam $b$, the mirror $m$ is then locked on the
reference mirror $r$ [see eq. (\ref{Equ_b2out})] and its residual motion no
longer depends on radiation pressure in the interferometer.\ It is related
only to error noises in the measurement by the control cavity.

The equivalent input noise of the interferometric measurement is obtained
from Eq. (\ref{Equ_Xsigfb}),
\begin{equation}
\Sigma _{{\rm sig}}=\frac{1}{4\xi _{{\rm a}}^{2}}+\frac{1}{4\xi _{{\rm b}%
}^{2}}+\frac{\hbar ^{2}\xi _{{\rm b}}^{2}}{\left| \Omega \right| ^{2}\left|
Z_{{\rm r}}\right| ^{2}}.  \label{Equ_Ssiginf}
\end{equation}
It only exhibits the phase noises of both beams and radiation-pressure
effects on the reference mirror $r$. The locking of mirror $m$ leads to a
transfer of quantum noise from the control measurement to the interferometric
one.\ The two last terms in Eq. (\ref{Equ_Ssiginf}) indeed correspond to the
equivalent input noise for the measurement by the control cavity [compare to
eq. (\ref{Equ_Ssigfree}) with $\xi _{{\rm a}}$ replaced by $\xi _{{\rm b}}$
and $Z_{{\rm m}}$ by $Z_{{\rm r}}$].

This quantum transfer is shown in curve {\it b} of Fig. \ref{Fig_Sensit},
obtained with a reference mirror of same mass as the mirror $m$ ($Z_{{\rm
r}}=Z_{{\rm m}}$) and with an optomechanical parameter $\xi _{{\rm b}}$ equal
to $\xi _{{\rm a}}/5$. At low frequency where radiation-pressure noise is
dominant, the mirror $m$ reproduces the motion of the reference mirror
induced by radiation pressure of beam $b$. Since the control measurement is
less sensitive ($\xi _{{\rm b}}<\xi _{{\rm a}}$), radiation-pressure effects
of beam $b$ are smaller than the ones of beam $a$ and the transfer of quantum
noise leads to a clear reduction of noise. As compared to the free
interferometer case (curve {\it a}), the equivalent input noise $\Sigma _{%
{\rm sig}}$ is reduced at low frequency by a factor $\left( \xi _{{\rm b}%
}/\xi _{{\rm a}}\right) ^{2}$ equal to $1/25$ in Fig. \ref{Fig_Sensit} and
which can be freely adjusted by changing the intensity of beam $b$.

For an infinite feedback gain this behavior is offset by the phase noise at
high frequency which is increased by the reverse factor.\ The frequency
dependence of the noise is thus similar to the free interferometer case
except that it is displaced along the standard quantum limit.

The loss of sensitivity at high frequency can easily be avoided by using a
frequency dependent feedback gain. For a finite gain the equivalent input
noise is given by
\begin{eqnarray}
\Sigma _{{\rm sig}} &=&\frac{1}{4\xi _{{\rm a}}^{2}}+\left| \frac{Z_{{\rm fb}%
}}{Z_{{\rm m}}+Z_{{\rm fb}}}\right| ^{2}\frac{1}{4\xi _{{\rm b}}^{2}}
\nonumber \\
&&+\frac{\hbar ^{2}}{\Omega ^{2}\left| Z_{{\rm m}}+Z_{{\rm fb}}\right| ^{2}}%
\left( \xi _{{\rm a}}^{2}+\left| \frac{Z_{{\rm r}}-Z_{{\rm fb}}}{Z_{{\rm r}}}%
\right| ^{2}\xi _{{\rm b}}^{2}\right)
\end{eqnarray}
The feedback gain $Z_{{\rm fb}}$ can be adapted in such a way that the
control is active at low frequency, whereas it plays no significant role at
high frequency.\ It is then possible to have an important reduction of
radiation-pressure effects by quantum transfer without losing the high
sensitivity of the interferometric measurement at frequencies where it is
limited by phase noise. Focusing on the case of two mirrors of equal masses (%
$Z_{{\rm r}}=Z_{{\rm m}}$), one can derive the optimum feedback gain $Z_{{\rm %
fb}}^{opt}$ which gives the minimum noise at every frequency,
\begin{eqnarray}
Z_{{\rm fb}}^{opt} &=&Z_{{\rm m}}\left( 1+\frac{\xi _{{\rm a}}^{2}}{2\xi _{%
{\rm b}}^{2}}\right) \frac{1}{1+\left( \Omega /\Omega _{{\rm fb}%
}\right) ^{4}},  \label{Equ_Zopt} \\
\Omega _{{\rm fb}}^{2} &=&\sqrt{2}\frac{2\hbar \xi _{{\rm b}}^{2}}{M_{%
{\rm m}}}.
\end{eqnarray}
$Z_{{\rm fb}}^{opt}$ is large at low frequency and quickly decreases for
frequencies larger than $\Omega _{{\rm fb}}$. Apart from a factor $%
\sqrt[4]{2}$, this cutoff frequency corresponds to the {\sc sql} frequency
of the control cavity [eq. (\ref{Equ_Osql}) with $\xi _{{\rm a}}$ replaced
by $\xi _{{\rm b}}$].

The resulting noise $\Sigma _{{\rm sig}}$ is shown in curve {\it c} of Fig.\
\ref{Fig_Sensit}.\ It exhibits a very clear reduction of radiation pressure
effects while the limitation by phase noise is identical to the case of a
free interferometer. The controlled interferometer is actually equivalent to
a free interferometer with a frequency dependent optomechanical coupling
equal to $\xi _{{\rm a}}$ at high frequency and to the much smaller value $%
\xi _{{\rm b}}$ at low frequency. It also compares with a free
interferometer whose mirror mass is increased by a factor $\left( \xi _{{\rm %
a}}/\xi _{{\rm b}}\right) ^{2}$ that is $25$ times the weight in the case of
Fig.\ \ref{Fig_Sensit}.

In the intermediate frequency range between the {\sc sql} frequencies of the
control cavity and of the interferometer, the noise stays near the standard
quantum limit and even goes down below. In contrast with the free
interferometer for which the standard quantum limit is reached at only one
frequency, the sensitivity of the controlled interferometer is very close to
the optimum in a large frequency domain between the two {\sc sql}
frequencies.

Finally, note that for a complete interferometer this local control has to be
applied to each sensitive mirror of the interferometer, namely, the four
mirrors of the Fabry-Perot cavities. The control, however, requires a less
sensitive measurement than the interferometer itself, and its implementation
seems easy to achieve with currently available technology\cite
{Rempe92,Cohadon99}.\ Taking, for example, the parameters of the V{\sc irgo}
interferometer ($15$ $kW$ light power in each Fabry-Perot arms with a global
finesse of $600$)\cite{Bradaschia90}, the optomechanical parameter $\xi _{%
{\rm b}}$ of Fig.\ \ref{Fig_Sensit} would correspond to a control cavity of
finesse $10^{4}$ with an incident light power of $5$ $mW$ only.

In a practical implementation the optimum feedback gain has also to be
approximated by an electronic transfer function. Although one can use the
powerful methods already developed for servo controls in gravitational-wave
interferometers, this approximation is not to be stringent since a simple
first-order low-pass filter is sufficient to observe the reduction of
radiation-pressure noise.

In conclusion, a local control of mirror motion by an optomechanical sensor
and a feedback loop allows one to efficiently reduce the radiation-pressure
effects in an interferometric measurement. The low frequency sensitivity is
improved without alteration in the high frequency domain, thus increasing the
interferometer bandwidth. This result shows that active control is a powerful
technique to reduce quantum noise. As compared to other methods, an essential
characteristic of this control is to be decoupled from other optimizations of
the interferometer. Its local implementation does not induce any additional
constraint on the interferometer parameters.

\end{document}